\documentclass[aps,prl,preprint,tightenlines,superscriptaddress,showpacs,byrevtex]{revtex4}
\usepackage{graphicx} 
\usepackage{dcolumn}  

\graphicspath{{ps}}

\newcommand{\leplep}{\ell^{+}\ell^{-}}
\newcommand{\jp}{J/\psi}
\newcommand{\psip}{\psi '}
\newcommand{\jpsi}{J/\psi}

\newcommand{\pipi}{\pi^{+}\pi^{-}}

\newcommand{\rt}{\rightarrow}
\newcommand{\etal}{\em et al.}

\begin{document}


\preprint{\vbox{ \hbox{   }
                 \hbox{BELLE-CONF-0439}
                 \hbox{ICHEP04 8-0685} 
}}

\title{ \quad\\[0.5cm] Properties of the X(3872) at Belle }

\affiliation{Aomori University, Aomori}
\affiliation{Budker Institute of Nuclear Physics, Novosibirsk}
\affiliation{Chiba University, Chiba}
\affiliation{Chonnam National University, Kwangju}
\affiliation{Chuo University, Tokyo}
\affiliation{University of Cincinnati, Cincinnati, Ohio 45221}
\affiliation{University of Frankfurt, Frankfurt}
\affiliation{Gyeongsang National University, Chinju}
\affiliation{University of Hawaii, Honolulu, Hawaii 96822}
\affiliation{High Energy Accelerator Research Organization (KEK), Tsukuba}
\affiliation{Hiroshima Institute of Technology, Hiroshima}
\affiliation{Institute of High Energy Physics, Chinese Academy of Sciences, Beijing}
\affiliation{Institute of High Energy Physics, Vienna}
\affiliation{Institute for Theoretical and Experimental Physics, Moscow}
\affiliation{J. Stefan Institute, Ljubljana}
\affiliation{Kanagawa University, Yokohama}
\affiliation{Korea University, Seoul}
\affiliation{Kyoto University, Kyoto}
\affiliation{Kyungpook National University, Taegu}
\affiliation{Swiss Federal Institute of Technology of Lausanne, EPFL, Lausanne}
\affiliation{University of Ljubljana, Ljubljana}
\affiliation{University of Maribor, Maribor}
\affiliation{University of Melbourne, Victoria}
\affiliation{Nagoya University, Nagoya}
\affiliation{Nara Women's University, Nara}
\affiliation{National Central University, Chung-li}
\affiliation{National Kaohsiung Normal University, Kaohsiung}
\affiliation{National United University, Miao Li}
\affiliation{Department of Physics, National Taiwan University, Taipei}
\affiliation{H. Niewodniczanski Institute of Nuclear Physics, Krakow}
\affiliation{Nihon Dental College, Niigata}
\affiliation{Niigata University, Niigata}
\affiliation{Osaka City University, Osaka}
\affiliation{Osaka University, Osaka}
\affiliation{Panjab University, Chandigarh}
\affiliation{Peking University, Beijing}
\affiliation{Princeton University, Princeton, New Jersey 08545}
\affiliation{RIKEN BNL Research Center, Upton, New York 11973}
\affiliation{Saga University, Saga}
\affiliation{University of Science and Technology of China, Hefei}
\affiliation{Seoul National University, Seoul}
\affiliation{Sungkyunkwan University, Suwon}
\affiliation{University of Sydney, Sydney NSW}
\affiliation{Tata Institute of Fundamental Research, Bombay}
\affiliation{Toho University, Funabashi}
\affiliation{Tohoku Gakuin University, Tagajo}
\affiliation{Tohoku University, Sendai}
\affiliation{Department of Physics, University of Tokyo, Tokyo}
\affiliation{Tokyo Institute of Technology, Tokyo}
\affiliation{Tokyo Metropolitan University, Tokyo}
\affiliation{Tokyo University of Agriculture and Technology, Tokyo}
\affiliation{Toyama National College of Maritime Technology, Toyama}
\affiliation{University of Tsukuba, Tsukuba}
\affiliation{Utkal University, Bhubaneswer}
\affiliation{Virginia Polytechnic Institute and State University, Blacksburg, Virginia 24061}
\affiliation{Yonsei University, Seoul}
  \author{K.~Abe}\affiliation{High Energy Accelerator Research Organization (KEK), Tsukuba} 
  \author{K.~Abe}\affiliation{Tohoku Gakuin University, Tagajo} 
  \author{N.~Abe}\affiliation{Tokyo Institute of Technology, Tokyo} 
  \author{I.~Adachi}\affiliation{High Energy Accelerator Research Organization (KEK), Tsukuba} 
  \author{H.~Aihara}\affiliation{Department of Physics, University of Tokyo, Tokyo} 
  \author{M.~Akatsu}\affiliation{Nagoya University, Nagoya} 
  \author{Y.~Asano}\affiliation{University of Tsukuba, Tsukuba} 
  \author{T.~Aso}\affiliation{Toyama National College of Maritime Technology, Toyama} 
  \author{V.~Aulchenko}\affiliation{Budker Institute of Nuclear Physics, Novosibirsk} 
  \author{T.~Aushev}\affiliation{Institute for Theoretical and Experimental Physics, Moscow} 
% ...
\collaboration{The Belle Collaboration}

\collaboration{Belle Collaboration}
\noaffiliation

\begin{abstract}
We report recent results on the properties of the $X(3872)$ produced
via the $B^{+} \rt K^{+} X(3872)$ decay process. 
We observe decays $X\rt\pipi\pi^0\jp$ where the $3\pi$
invariant masses cluster near the upper kinematic boundary
suggesting that they originate from sub-threshold decays to virtual
$\omega(782)$ mesons.  This is consistent
with expectations for a $D\bar{D^*}$ bound state
interpretation for the $X(3872)$.  In addition, we constrain 
the possible  charmonium-state assignments for this particles.
Results are obtained from a $253\,{\rm fb}^{-1}$ data sample that contains 274
million $B\bar{B}$ pairs that was collected 
near the $\Upsilon(4S)$ resonance
with the Belle detector at the KEKB asymmetric energy $e^+ e^-$
collider.
\end{abstract}


\maketitle

\tighten

{\renewcommand{\thefootnote}{\fnsymbol{footnote}}}
\setcounter{footnote}{0}

\section{Introduction}
A first step in understanding the $X(3872)$ particle that was recently
discovered by Belle~\cite{s-s} is to determine its $J^{PC}$ 
quantum numbers.  Here, we survey possible assignments and properties 
of the most likely candidates and contrast these with recent
experimental measurements.

Although the $X(3872)$ is above the $D\bar{D}$ mass threshold, its
width is narrow, $\Gamma < 2.3$~MeV~\cite{s-s}, and decays to $D\bar{D}$ are
not seen~\cite{chistov}.  This suggests that $D\bar{D}$ decays are
forbidden.  We restrict our considerations
to $0^{++}$ and $1^{--}$ $\pipi$ systems~\cite{close} and scenarios 
where the relative 
orbital angular momentum of the $\pipi$ and $\jp$ is $L \le 3$.
We concentrate on possible charmonium assignments,
and only those where decays to 
$D\overline{D}$ are forbidden or expected to be strongly suppressed.
For the case of a $0^{++}$ dipion, there are three charmonium 
possibilities:
the $h_c ^{'} (2^1P_1)$
and two triplet D-wave states, the $\psi_2 (^3D_2)$ and 
$\psi_3 (^3D_3)$.
For the $1^{--}$ dipion case, there are also three possibilities: the 
$\eta_{c} ^{''}$, the $\chi_{c1} ^{'}$ and the $\eta_{c2} (^1D_2)$.
For these assignments, the $\pipi \jp$ decay would violate isospin and 
should be strongly suppressed.

The Belle experiment observes $B$ mesons produced by the KEKB
asymmetric energy $e^+e^-$ collider~\cite{KEKB}.
KEKB operates at the $\Upsilon(4S)$ resonance
($\sqrt{s}=10.58$~GeV) with
a peak luminosity of $1.39\times 10^{34}~{\rm cm}^{-2}{\rm s}^{-1}$. 
At the $\Upsilon(4S)$, $B\bar{B}$ meson pairs are
produced with no accompanying particles. As a result, 
the $B$ mesons have a total center-of-mass system (cms) 
energy that is equal to $E_{beam}$, the cms beam energy.  
We identify $B$ mesons using the beam-constrained mass 
$M_{bc}=\sqrt{E_{beam}^2 - p_B^2}$ and the energy difference
$\Delta E = E_{beam} - E_B$, where $p_B$ is the vector sum of the 
cms momenta of the $B$ meson decay products and $E_B$ is their
cms energy sum.  The experimental resolution of $M_{bc}$ is 
approximately 3~MeV; that for $\Delta E$ is typically 11~MeV
for all-charged-particle final states.  For final states
with $\gamma$'s or $\pi^0$'s, the $\Delta E$ resolution becomes
broader and somewhat skewed to negative values
due to energy leakage out of the back of electromagnetic
calorimeter.  The Belle detector is described in ref.~\cite{Belle}. 

\section{Search for $X(3872)\rt \gamma \chi_{c2}$ ($\chi_{c1}$)}
The Wigner-Eckart theorem says that 
the widths for $\psi_{2} \rt \pipi \jp$ and $\psi_3 \rt \pipi \jp$ should 
both equal $\Gamma (\psi (3770) \rt \pipi \jp)$.
This has been recently measured by BESII~\cite{bes2} and CLEO-c~\cite{cleoc}
to be $80 \pm 32 \pm 21$ keV and $\le 55$ keV (90$\%$ CL), respectively.
The results are in some contradiction with each other. For the following
discussion we conservatively assume an upper limit derived from
the larger BES number of $\Gamma (\psi (3770) \rt \pipi \jp) < $ 130 keV.

Calculations of the $\gamma \chi_{c1}$ width for an $M=3872$~MeV $\psi_2 $
range from 207~keV~\cite{eichten} to
360~keV~\cite{barnes}.
The 90\% CL upper limit of 
\begin{eqnarray}
 \frac {\Gamma (X \rt \gamma \chi_{c1})}{\Gamma (X \rt \pipi \jp)} < 0.89
\end{eqnarray}
that was reported in ref.~[1] contradicts these expectations for the 
$\psi_2$.

Barnes and Godfrey~\cite{barnes} observe that 
although $\psi_{3} \rt D \overline{D}$
is allowed for a 3872 MeV $\psi_{3}$, this mode is suppressed by an $L=3$ 
centrifugal 
barrier and the total $\psi_{3}$ width may be less than the $\Gamma<2.3$ 
MeV experimental upper limit.
These authors, and also Eichten, Lane and Quigg~\cite{eichten}, propose 
the $\psi_{3}$ as a charmonium candidate for the $X(3872)$.

For an $M=3872$~MeV $\psi_3$, the calculated $\gamma \chi_{c2}$ widths 
range from 299~keV\cite{eichten} to 370~keV~\cite{barnes}. 
Thus, the partial width for 
$\psi_{3} \rt \gamma \chi_{c2}$ is expected to be more than
twice that for $\psi_{3} \rt \pipi \jp$.
We performed a search for $X \rt \gamma \chi_{c2}$
that followed closely the procedure used for the $\gamma \chi_{c1}$ limit 
reported in ref.~[1].  We used a $140~{\rm fb}^{-1}$ data sample,
which contains 152 million $B\overline{B}$ pairs.
We require one of the $\gamma \jp$ combinations to 
satisfy 444~MeV $<(M_{\gamma \leplep} - M_{\leplep} ) <$ 469~MeV. 
The $M_{bc}$ and $\Delta E$ signal regions are $|M_{bc} - 5.28|<$ 0.009 GeV
and -0.04 $< \Delta E <$ 0.03~GeV.

\begin{figure}[htb]
\includegraphics[width=0.54\textwidth]{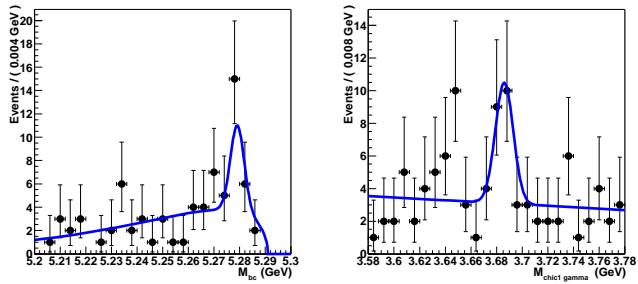}
\caption{Signal-band projections of $M_{bc}$ (left) and $M_{\gamma \chi_{c2}}$ 
(right) for events in the $\psi^{'}$ region with the 
results of the unbinned fit superimposed.}
\label{psip2gammachic2}
\end{figure}

We use the $B \rt K \psi^{'}; \psi^{'} \rt \gamma \chi_{c2}$ decay chain as a 
normalization reaction. The signal-band projections of $M_{bc}$ and 
$M_{\gamma \chi_{c2}}$ for the $\psi^{'}$ region are shown 
in Figs.~\ref{psip2gammachic2} (left) and (right), respectively, together 
with curves that show the results of the fit. The fitted signal yield is
$18.3 \pm 5.2$ events, where, based on known branching fractions,
we expect $12\pm 3$ events.

\begin{figure}[htb]
\includegraphics[width=0.54\textwidth]{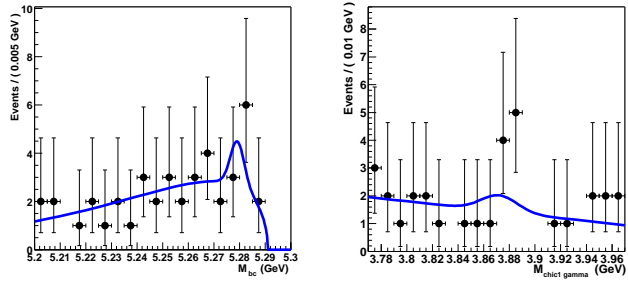}
\caption{Signal-band projections of $M_{bc}$ (left) and $M_{\gamma \chi_{c2}}$ 
(right) for events in the $X(3872)$ region with the 
results of the unbinned fit superimposed.}
\label{x2gammachic2}
\end{figure}

Figure~\ref{x2gammachic2} show the same projections for events 
in the $X(3872)$ mass region, where
there is no apparent signal. An unbinned fit produces a signal yield of 
$2.9 \pm 3.0 \pm 1.5$ events, where the first error is statistical and the 
second systematic. The latter is estimated by the changes that occur when 
the input parameters to the fit are varied over their allowed range of values.

The ratio of the $X \rt \gamma \chi_{c2}$ and the $X \rt \pipi \jp$ partial 
widths and its 90$\%$ CL upper
limit are
\begin{eqnarray}
 \frac {\Gamma (X \rt \gamma \chi_{c2})}{\Gamma (X \rt \pipi \jp)} 
= 0.42 \pm 0.45 \pm 0.23 < 1.1 ( 90\% CL),
\end{eqnarray}
where the second quoted error is the quadratic sum of the systematic 
uncertainties in acceptance, the branching fractions and variations in the 
$\gamma \chi_{c2}$ event yield for different fitting methods.

\section{Search for $X \rt \gamma \jp$}
The $\chi_{c1}^{'}$ is expected to be near 3968~MeV, well above the 
$D \overline{D}^{*}$ threshold, and its width is expected to be hundreds
of MeV\cite{eichten}.  If potential models are wrong and the
$\chi_{c1}^{'}$ is below the $D \overline{D}^{*}$ threshold
at 3872~MeV, it could conceivably be narrow and $\pipi \jp$ decays 
might be significant, even though these would violate isospin.
In this case, the $\gamma \psi^{'}$ and $\gamma \jp$ transitions would
be important and almost certainly have larger partial widths than that 
for the $\pipi \jp$ mode. We searched for the $X \rt \gamma \jp$ decay mode.

\begin{figure}[htb]
\includegraphics[width=0.54\textwidth]{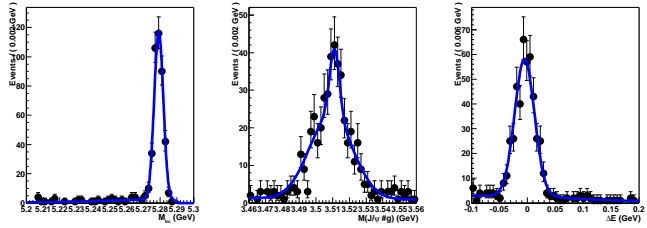}
\caption{ Signal-band projections of $M_{bc}$ (left), $M_{\gamma \jp}$ (center)
and $\Delta E$ (right) for events in the $\chi_{c1}$ region with the 
results of the unbinned fit superimposed.}
\label{chi12gammajpsi}
\end{figure}

\begin{figure}[htb]
\includegraphics[width=0.54\textwidth]{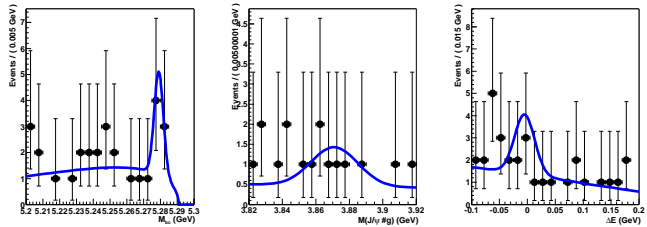}
\caption{
Signal-band projections of $M_{bc}$ (left), $M_{\gamma \jp}$ (center)
and $\Delta E$ (right) for events in the $X(3872)$ signal region with the 
results of the unbinned fit superimposed.}
\label{x2gammajpsi}
\end{figure}

We select $B^{+} \rt K^{+} \gamma \jp$ event candidates using 
the criteria given in ref.~[1]. 
The $B^{+} \rt K^{+} \gamma \jp$ channel is dominated by 
$B^{+} \rt K^{+} \chi_{c1}$; $\chi_{c1} \rt \gamma \jp$ decays and we use
this as a calibration reaction. 
We define a $\chi_{c1}$ window for $\gamma \jp$ masses within 20 MeV of the 
nominal $\chi_{c1}$ mass. Figure~\ref{chi12gammajpsi} shows the signal-band 
projections for $M_{bc}$ (left), 
$M_{\gamma \jp}$ (center) and $\Delta E$ (right) 
for events in the $\chi_{c1}$ region with the results of a 
three-dimensional unbinned fit superimposed. 
The fitted number of events is $470\pm 24$.

We define an $X \rt \gamma \jp$ signal region to be 
$|M(\gamma \jp)-3872~{\rm MeV}|
<26$ MeV. Figure~\ref{x2gammajpsi} shows the same 
projections for events in the X(3872) signal region. 
Here there is no strong evidence for a signal: the fit gives a
$2.2\sigma$ signal yield  of $7.7\pm 3.6$ events.
The resulting limit is 
\begin{eqnarray}
 \frac {\Gamma (X \rt \gamma \jp)}{\Gamma (X \rt \pipi \jp)} 
= 0.22 \pm 0.12 \pm 0.06 <0.40 ( 90\% CL),
\end{eqnarray}
where the second quoted error is systematic and includes uncertainties in 
acceptance, the branching fractions and variations in the $\gamma \jp$ 
event yield for different fitting methods.

\section{Helicity angle distribution for $1^{+-} h_{c}^{'}$}
\begin{figure}[htb]
\includegraphics[width=0.4\textwidth]{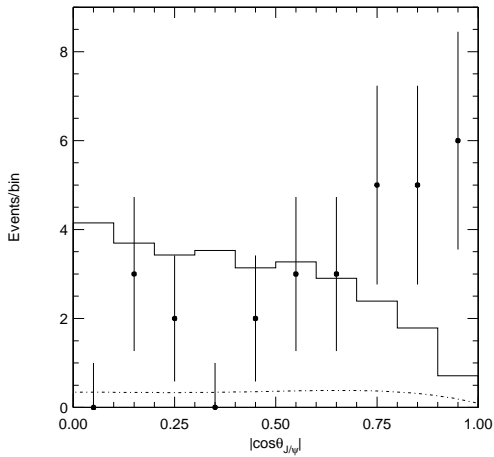}
\caption{The measured $| \cos\theta_{\jpsi} |$ distribution. 
The superimposed histogram is the normalized MC distribution
for the 1+- hypothesis. Here $\chi ^{2} / dof$ = 75/9.}
\label{helicity1+-}
\end{figure}

The $h_c'$ hypothesis makes reasonably specific predictions
for the $X\rt\pipi\jp$ decay angular distributions~\cite{suzuki}. 
We define $\theta_{\jp}$  as the angle between the $\jp$ and 
the negative of the $K^{+}$ momentum vectors in the $X(3872)$ rest frame
in the decay $B \rt X K; X \rt \pipi \jp$. 
The $|\cos\theta_{\jp}|$ distribution for
$X(3872)$ events with $m_{\pipi}>$ 0.65 GeV is shown as
data points in Fig.~\ref{helicity1+-}. The smooth dotted curve is
a polynomial represents sideband-determined backgrounds.

Figure~\ref{helicity1+-} shows a comparison of the measured 
$|cos \theta_{\jp}|$ distribution 
with a MC sample generated with a $J^{PC}=1^{+-}$ hypothesis. 
Here the expected $|\cos \theta_{\jp}|$ distribution has 
a $\sin^2 \theta_{\jpsi}$ dependence that goes to zero at
$\cos \theta_{\jp} = 1$, where the data tend to peak. 
This makes the overall $\chi^{2}$ quite poor, $\chi^{2} / dof$ is 75/9, 
and enables us to rule out the $1^{+-}(h_{c} ^{'})$
hypothesis for the $X(3872)$ with high confidence.

\section{Search for $X(3872)\rt\pi^0\pi^0\jp$}

The ratio ${\mathcal R}_0 = \Gamma(X\rt\pi^0\pi^0\jp)/\Gamma(X\rt\pipi\jp)$
measures the isospin of the dipion system~\cite{voloshin}. If isospin is 
conserved, $I=0$ corresponds to  ${\mathcal R}_0 = 1/2$; for $I=1$,
${\mathcal R}_0 =0$.  For the $\psip$, where the dipion system
is known to have $I=0$, this ratio is $0.60\pm0.05$~\cite{PDG}.
We searched for  $X\rt \pi^0\pi^0\jp$ decays
using a $253\,{\rm fb}^{-1}$ data sample that contains 274
million $B\bar{B}$ pairs.  We
use $B\rt K\psip$; $\psip\rt\pi^0\pi^0\jp$ as a calibration reaction.

We select $B\rt K \pi^0\pi^0\jp$ events using the
$\jp$ and charged kaon criteria given in ref.~[1].  For
neutral $B$s we use the standard Belle $K_S\rt\pipi$ criteria
and require the $\pipi$ invariant mass to be within $\pm 15$~MeV ($\simeq 3\sigma$) 
of $M_{K^0}$.  We identify $\pi^0$s as $\gamma\gamma$ pairs that
fit the $\pi^0\rt\gamma\gamma$ hypothesis with $\chi^2<6$.  We further
require the energy asymmetry 
$|(E_{\gamma_1}-E_{\gamma_2})/(E_{\gamma_1}+E_{\gamma_2})|<0.9$ and
the $\pi^0$ cms momentum to be greater than 150~MeV.  For cases
where there are more than two $\pi^0$ candidates in the
$|\Delta E|<0.2$~GeV and $M_{bc}>5.2$~GeV selection region,
we take the combination with the minimum value of
$(M_{\gamma\gamma}-M_{\pi^0})^2/\sigma_{\pi^0}^2 + |\Delta E|^2/\sigma_{\Delta E}^2$.

\begin{figure}[htb]
\includegraphics[width=0.7\textwidth]{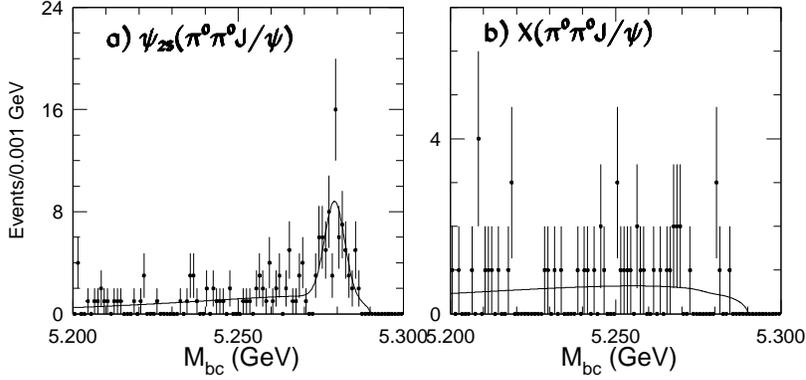} 
\caption{{\bf (a)} The $M_{bc}$ distribution for candidate
$B\rt K\psip$; $\psip\rt\pi^0\pi^0\jp$ decays.  The curve is the
result of the fit described in the text. {\bf (b)} The
$M_{bc}$ distribution for $B\rt K X$; $X\rt\pi^0\pi^0\jp$ decays.}
\label{fig:pi0pi0_mbc}
\end{figure}

Figure~\ref{fig:pi0pi0_mbc}(a) shows the $M_{bc}$ distribution for
events in the $-0.06~{\rm GeV}<\Delta E< 0.03~{\rm GeV}$ signal
region with $M_{\pi^0\pi^0\jp}$ within $\pm 15$~MeV of $M_{\psip}$.
Here we have also required $M_{\pi^0\pi^0}>390$~MeV.  We
fit this distribution with a Gaussian function to represent
the signal and a smoothed threshold function to represent
the background.  The signal yield from the fit is $55\pm 10$ events.

Figure~\ref{fig:pi0pi0_mbc}(b) shows the corresponding
$M_{bc}$ distribution for events
with $M_{\pi^0\pi^0\jp}$ within $\pm 15$~MeV of 
3872~MeV.  We have applied an additional requirement 
$M_{\pi^0\pi^0}>570$~MeV.  This dipion mass restriction has 
$\simeq$100\% acceptance
for $X\rt\pipi\jp$ decays and we assume a similar efficiency
for $\pi^0\pi^0\jp$.  Here there is no evident signal; a fit
gives a signal yield of $0.2\pm 2.6$ events.

We compare these signal yields to corresponding results for
$\psip$ and $X(3872)$ decays to $\pipi\jp$ and determine
the 90\% CL upper limit:
\begin{equation}
\frac{\Gamma(X\rt\pi^0\pi^0\jp)}{\Gamma(X\rt\pipi\jp)} < 1.3
\frac{\Gamma(\psip\rt\pi^0\pi^0\jp)}{\Gamma(\psip\rt\pipi\jp)}.
\end{equation}
By quoting our result this way, many systematic effects, including
the influence of the dipion mass requirements, cancel out.  Unfortunately,
with the present data sample, the limit is not stringent enough
to distinguish between the $I=0$ and $I=1$ hypotheses.

\section{Observation of $X(3872)\rt\omega\jp$}

In the context of an analysis of the possibility that the $X(3872)$
might be a weakly bound $J^{PC}=1^{++}$ $D^0\bar{D^{0*}}$ molecular state,
Swanson~\cite{swanson} developed a specific model for
the $X(3872)$ as a $D^0\bar{D^{0*}}$ hadronic resonance 
with an important admixture of $\omega\jp$.  He
finds that although the $X$ is 7.5~MeV below
the $M_{\jp}+M_{\omega}$ mass threshold, decays
that proceed via virtual $\omega$ mesons are important;
the signature being $B\rt K X$; $X\rt \pipi\pi^0\jp$, where
all the events have $\pipi\pi^0$ invariant mass values near
the kinematic upper limit. He predicts that these decays
should occur at about half the rate for $X\rt\pipi\jp$.
We searched for $B\rt K X$; $X\rt \pipi\pi^0\jp$ decays
in a 274 million $B\bar{B}$ event sample.

We select events with the same kaon, charged pion and
$\jp$ requirements as used in ref.~[1].  An additional
$\pi^0\rt\gamma\gamma$ candidate that
satisfies the same criteria listed above for the
$X\rt\pi^0\pi^0\jp$ search is required.  
The $\pi^0$ cms momentum
is required to be above 180~MeV.  We select
events where $M(\pipi\pi^0\jp)$ is within
$\pm 12$~MeV ($2\sigma$) of 3872~MeV.  The
$M_{bc}$ and $\Delta E$ signal regions are
$|M_{bc}-M_B|<7.5$~MeV and $|\Delta E|<30$~MeV.

\begin{figure}[htb]
\includegraphics[width=0.7\textwidth]{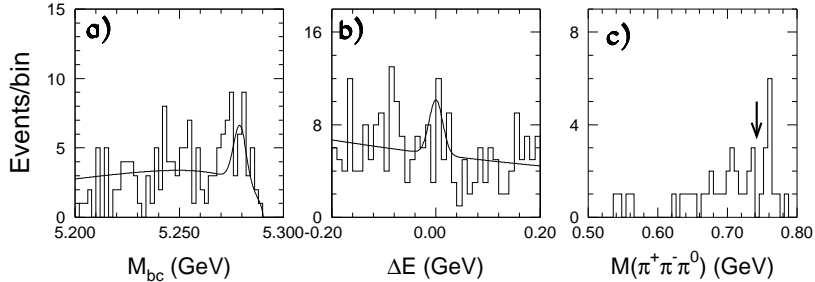} 
\caption{{\bf (a)} The $M_{bc}$ and
{\bf (b)} $\Delta E$ distributions for candidate
$B\rt K\pipi\pi^0 \jp$ decays.  The curves are the
result of the fit described in the text. {\bf (c)} The
$M(\pipi\pi^0)$ distribution for events in the
$M_{bc}$-$\Delta E$ signal region.}
\label{fig:x_2_3piJpsi}
\end{figure}

Figure~\ref{fig:x_2_3piJpsi}(a) shows the $M_{bc}$
projection for events in the $\Delta E$ signal 
region; Fig.~\ref{fig:x_2_3piJpsi}(b) shows the
corresponding $\Delta E$ projection.  There is
some indication of a $B$ meson signal: a simultaneous 
fit to the two distributions gives a signal yield
of $15.4\pm 6.3$ events, with a $S/B=0.8$. 
Figure~\ref{fig:x_2_3piJpsi}(c) shows the 
$\pipi\pi^0$ invariant mass for events in the
$M_{bc}$-$\Delta E$ signal region, where
a peak is evident at high masses.

\begin{figure}[htb]
\includegraphics[width=0.7\textwidth]{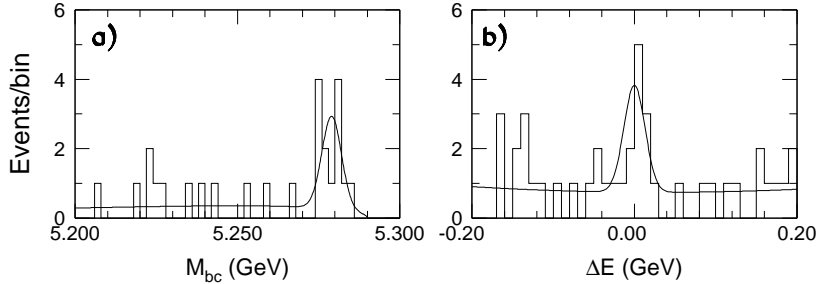} 
\caption{{\bf (a)} The $M_{bc}$ and
{\bf (b)} $\Delta E$ distributions for candidate
$B\rt K\pipi\pi^0 \jp$ decays with $M(\pipi\pi^0)>0.75$~GeV.  
The curves are the result of the fit described in the text.}
\label{fig:omega_jpsi_2box}
\end{figure}
 
Figures~\ref{fig:omega_jpsi_2box}(a) and (b) show the
$M_{bc}$ and $\Delta E$ projections for events with
$M(\pipi\pi^0)>0.75$~GeV (indicated by the arrow
in Fig.~\ref{fig:x_2_3piJpsi}(c)).   Here a $B$ meson signal
is evident on a small background.  The signal yield
from a simultaneous fit is $10.0 \pm 3.6$ events
and $S/B=5$. 
The statistical significance
of the signal, determined from
$\sqrt{-2\ln({\mathcal L}_0/{\mathcal L}_{\rm max})}$,
where ${\mathcal L}_{\rm max}$ and ${\mathcal L}_0$ are the likelihood
values for the best-fit and for zero-signal-yield, respectively, 
is $5.8\sigma$.   This is the first obervation of an $X(3872)$ decay
mode other than $\pipi\jp$.

We attribute all of the signal events with $\pipi\pi^0$ invariant
mass greater than $0.75$~GeV to $B\rt\omega\jp$ and 
compute the ratio of $\omega\jp$ and $\pipi\jp$ partial widths
by comparing this to the number of  $X\rt\pipi\jp$
in the sample data sample, corrected by the relative detection efficiencies:
\begin{equation}
\frac{\Gamma(X\rt\omega\jp)}{\Gamma(X\rt\pipi\jp)} = 0.8\pm0.3{\rm (stat)}\pm0.1{\rm (syst)},
\end{equation}
where the systematic error reflects the uncertainty in the 
relative acceptance.

The properties of the $X\rt \pipi\pi^0\jp$ decays are in
good agreement with expectations of ref.~\cite{swanson}. 
The $\pipi\pi^0$ invariant masses cluster near the upper
kinematic limit, and its measured strength 
is consistent with being ``roughly 1/2'' that for the $\pipi\jpsi$
mode.

\section{Summary}

The observation of $X(3872)\rt\pipi\pi^0\jp$ decays with
properties consistent with expectations for $X\rt\omega\jp$
provides strong support for the $D\bar{D^*}$ molecular state
interpretation for the $X(3872)$.  Both the observed
$3\pi$ mass distribution and the decay srength relative
to $\pipi\jp$ agree with predictions by Swanson~\cite{swanson}.

Moreover, none of the six possible charmonium candidate states comfortably fit the
measured properties.
The $90\%$ CL branching fraction upper limit for $\it B (X(3872)
\rt \gamma \chi_{c2}$) decay is 1.1 times that for $\pipi \jp$. 
This conflicts with theoretical expectations for the case where the X(3872)
is the $3^{--} \psi_{3}$.

The possibility that the $X(3872)$ is the $1^{++} \chi_{c1}^{'}$
charmonium state is made improbable by the limit 
${\cal B}(X \rt \gamma \jp) < 0.4 {\cal B} (X \rt \pipi \jp)$.
The former would be an allowed E1 transition 
with an expected width of $\Gamma_{\gamma \jp} \sim 10$~keV\cite{barnes}.
The latter would be an isospin-violating transition; other isospin 
violating 
transitions in the charmonium system have widths that are less than 1 keV.

An analysis of the $\theta_{\jp}$ helicity angle distribution 
eliminates the $1^{+-}(h_{c}^{'})$ hypothesis with a high degree of
confidence.

The $0^{-+}(\eta_{c}^{''})$ mass differs from that of the 
$\psi(3S)$ by hyperfine splitting and can be reliably expected to be about 
50 MeV (or less) below that of the $\psi(3S)$, which is at 4030~MeV.
Moreover, even if it were as low as 3872 MeV, the width is expected to be
some 10's of MeV, similar to that of the $\eta_{c}$ and wider than the
2.3~MeV upper limit for the X(3872). 
For $2^{-+}(\eta_{c2})$, the $\eta_{c2} \rt \pipi \eta_{c}$ and $\gamma h_{c}$
decays are allowed and expected to have widths in the range of 100's of 
keV\cite{barnes}, and much larger than that for the isospin-violating 
$\pipi \jp$ mode.  If the X(3872) were the $\eta_{c2}$, 
the total exclusive
branching fraction for the $B^{+} \rt K^{+} \eta_{c2}$ decay, 
which is non-factorizable and suppressed by an $L=2$ barrier, 
would be anomalously large.

\section*{Acknowledgments}
We thank the KEKB group for the excellent operation of the
accelerator, the KEK Cryogenics group for the efficient
operation of the solenoid, and the KEK computer group and
the National Institute of Informatics for valuable computing
and Super-SINET network support. We acknowledge support from
the Ministry of Education, Culture, Sports, Science, and
Technology of Japan and the Japan Society for the Promotion
of Science; the Australian Research Council and the
Australian Department of Education, Science and Training;
the National Science Foundation of China under contract
No.~10175071; the Department of Science and Technology of
India; the BK21 program of the Ministry of Education of
Korea and the CHEP SRC program of the Korea Science and
Engineering Foundation; the Polish State Committee for
Scientific Research under contract No.~2P03B 01324; the
Ministry of Science and Technology of the Russian
Federation; the Ministry of Education, Science and Sport of
the Republic of Slovenia; the National Science Council and
the Ministry of Education of Taiwan; and the U.S.\
Department of Energy.

\end{document}